# Modeling of a chlorine high-density plasma submitted to a static magnetic field

J.S. Poirier[1], L. Stafford[1], J. Margot[1], F. Vidal[2], K. Giroux[1], A. Quintal-Léonard[1] & M. Chaker[2]

[1] *Groupe de physique des plasmas, Université de Montréal, Montréal, Québec Canada*
[2] *INRS-Énergie, Matériaux et Télécommunications, Varennes, Québec, Canada*

**Abstract**
This paper extends the results of a recently developed one-dimensional model aiming to describe the characteristics of a magnetized chlorine high-density plasma. In this work, the dependence of the plasma characteristics on the magnetic field intensity is investigated. It is shown that the dissociation degree and the relative weight of the various charged species is strongly influenced by the magnetic field when the gas pressure is low enough. In contrast, at higher pressure, the plasma is essentially composed of negative ions, and molecular neutrals and ions, independently of the field intensity. It is further demonstrated that diffusion needs to be considered in order to correctly predict the plasma behavior.

**1. Introduction**
In a previous work [1], we have presented preliminary results of a 1-dimensional fluid model aiming to describe the characteristics of a magnetized chlorine high-density plasma in a pressure regime suitable for nanometer etching applications. This model allows to predict the dependence of various plasma characteristics, such as the concentration of charged and neutral species, on the experimental conditions. Previous results have shown that the model correctly predicts the dependence of the Cl and $Cl_2$ densities on the gas pressure as well as that of $Cl^+$ ions and electrons. In the present paper, we extend this work by examining the influence of a static magnetic field on the plasma characteristics. In particular, the densities of electrons, Cl, $Cl_2$, $Cl^+$, $Cl_2^+$ and $Cl^-$ are calculated. The results are discussed by putting the emphasis on the relative importance of volume reactions with respect to diffusion in the particle balance of charged species.

**2. Overview of the model**
The model was discussed in more details in a previous publication [1]. In brief, it is based on fluid equations, in which the particle balance equations for electrons, Cl, $Cl_2$, $Cl^+$, $Cl_2^+$ and $Cl^-$ are solved together with the corresponding flux equations and the energy balance equation. For simplicity, the model is isothermal, which means that power deposition is considered as spatially uniform. The electronegative nature of chlorine requires to treat the plasma sheath in detail by solving Poisson's equation. In these conditions, the set of equations to be solved takes the following form

$$\frac{\partial n_k}{\partial t} + \nabla \cdot \mathbf{J}_k = S_k, \qquad (1)$$

$$\mathbf{J}_k = \pm n_k \mu_k \mathbf{E} - D_k \nabla n_k, \qquad (2)$$

$$P_V = \bar{n}_e \Theta, \qquad (3)$$

$$\nabla \cdot \mathbf{E} = e\left(-n_e - n_{Cl^-} + n_{Cl_2^+} + n_{Cl^+}\right)/\varepsilon_0, \qquad (4)$$

where k=1-6 indicates one of the 6 species considered, $n_k$ being the density of the species k, $\mathbf{J}_k$ its flux, $S_k$ its source term (volume creation minus volume losses), $D_k$ and $\mu_k$ its free diffusion coefficient and mobility, $\mathbf{E}$ the space-charge electric field, $P_V$ the average power absorbed by the plasma per volume unit and $\Theta$ the average power dissipated per electron [2]. It is calculated by summing the power losses due to all electron-neutral collision processes in the volume, [3] neglecting the power losses to the wall.

To avoid the problem of solving a 2-D model, the axial contribution of the flux in Eq. (1) is treated as an adjustable constant of the form $-(D_a/\Lambda^2)n_k$, where $D_a$ is taken as the ambipolar diffusion coefficient and $\Lambda$ an axial diffusion length. As will be seen, although being oversimplified, this approach allows to reproduce the dependence of the plasma characteristics upon the magnetic field intensity.

The equation system is solved using usual boundary conditions

$$J_{kr}(0) = 0; \quad n_i(R+\delta) = 0; \quad J_{er}(R) = \frac{1}{4}\sqrt{\frac{8T_e}{\pi m_e}}n_e(R); \quad J_{Clr}(R) = \frac{\gamma}{4}\sqrt{\frac{8T_n}{\pi m_{Cl}}}n_{Cl}(R); \quad J_{Cl_2 r}(R) = -\frac{1}{2}J_{Clr}(R), \quad (5)$$

where R is the vessel radius, $n_i$ the total positive ion density, $T_e$ and $T_n$ the electron and neutral temperatures, and $\gamma$ the surface recombination coefficient characterizing the interaction of Cl atoms with the wall. In our experimental conditions, the value of $\gamma$ that provides the best agreement with the experimental data was found to be 0.02 [1].

The model is applied to the case of a bench-test high-density plasma produced by a 190 MHz electromagnetic surface wave [4]. The plasma can be confined by a static magnetic field, whose strength $B_0$ can be varied from 0 to about 1 kG. Note that the parameters used for calculations were chosen to match the most usual experimental conditions achieved in this reactor : $P_V = 2.5$ mW/cm$^3$ (250 W absorbed power), $T_n = 300$ K, $\Lambda = 140$ cm. As our model is isothermal, the details of the radial plasma structure are not adequately described. Consequently, it is more suitable to compare model and experiment by using cross-section averaged values rather than local values.

## 3. Results of the model and comparison with experiments
### 3.1 Characteristics of neutral species
Figure 1 shows the dissociation degree $\tau_d$ of the Cl$_2$ molecules as a function of the initial gas pressure for different values of the magnetic field intensity. For comparison (and testing of the model validity), experimental data obtained at $B_0$=600 G, using the Cl$_2$ molecule actinometry technique proposed by Donnelly [5], are also shown [6]. For all $B_0$ values, $\tau_d$ is observed to decrease with increasing pressure, in agreement with the available experimental data. However, the variation of $\tau_d$ with pressure is more important when $B_0$ is stronger. Thus, at low pressure, the gas phase is mostly populated by Cl atoms while at higher pressure, the molecular concentration becomes significant. In contrast, at higher pressure, the gas phase is almost independent of the magnetic field and is dominated by Cl$_2$ molecules.

We have examined in more details the dependence of the dissociation degree on the magnetic field intensity for the two extreme pressure values investigated in Fig. 1, i.e. 1 and 10 mTorr. The results are shown in Fig. 2. At 1 mTorr, $\tau_d$ increases very steeply with $B_0$ and reaches a plateau as soon as $B_0$ gets higher than about 50 G. In this case, it is clear that applying a weak magnetic field presents a significant advantage to enhance the Cl fraction in the plasma. In contrast, at 10 mTorr, $\tau_d$ is almost independent of $B_0$ and the plasma remains essentially molecular. In these conditions, there is no more benefit to apply a magnetic field to the discharge.

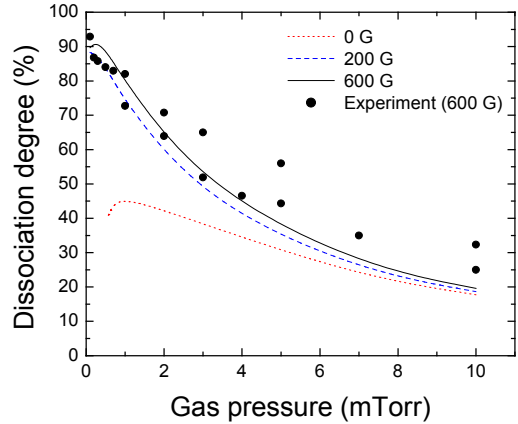

**Figure 1:** Dissociation degree of $Cl_2$ as a function of initial gas pressure for different values of the magnetic field intensity. Experimental data are taken from [6].

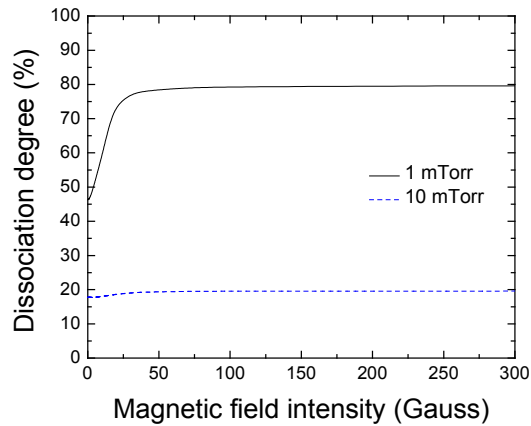

**Figure 2 :** Dissociation degree of $Cl_2$ as a function of magnetic field intensity for two values of the gas pressure.

### *3.2 Characteristics of charged species*
Figure 3 presents the density of electrons, $Cl^-$, $Cl^+$, and $Cl_2^+$ as a function of the magnetic field intensity at 1 mTorr. As can be seen, all these densities increase with $B_0$. This can be attributed to the decrease of the charged species losses in the radial direction (i.e. perpendicularly to the magnetic field). Indeed, above some $B_0$ value, diffusion across the field becomes so small that the process is then essentially governed by the axial motion and is therefore independent of $B_0$. A similar behavior for the electron density was previously observed for a pure argon plasma [7].

On the other hand, Fig. 3 shows that the main negative charge carriers are the negative ions at low magnetic fields and the electrons at higher $B_0$. This can be easily understood from the observations of Fig. 2. Indeed, negative ions are mainly created through dissociative attachment, $Cl_2 + e^- \rightarrow Cl^- + Cl$ [1]. As a result, their relative importance in the plasma is directly related to the $Cl_2$ concentration; the higher this relative concentration, the larger the population of $Cl^-$ negative ions with respect to the electron population. Similar observations can be made when comparing $Cl_2^+$ and $Cl^+$ densities. The molecular ion dominates the ion population for molecular plasmas while the opposite occurs for atomic plasmas. As creation is dominated by direct ionization of Cl atoms and $Cl_2$ molecules [1], the larger is the molecular content of the plasma, the higher is the resulting $Cl_2^+$ density.

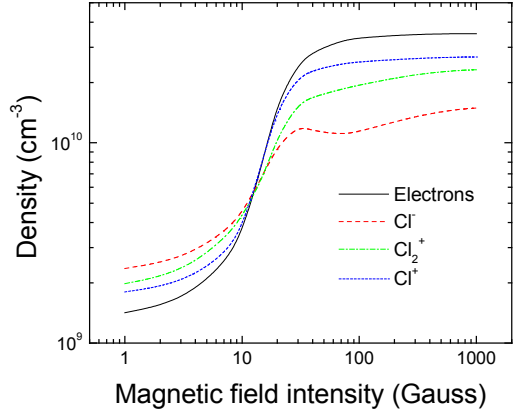

**Figure 3:** Density of charged particles as a function of the magnetic field intensity for an initial gas pressure of 1 mTorr.

As illustrated in Fig. 4, at 10 mTorr, the plasma exhibits a quite different behavior. In this case, the density of charged species is practically independent of $B_0$. This clearly emphasizes that the magnetic confinement does not influence the plasma any more, in agreement with our observations on the dissociation degree. Figure 4 further indicates that $Cl^-$ and $Cl_2^+$ are the main negative and positive charge carriers, the electron and $Cl^+$ densities being more than one order of magnitude smaller. As mentioned earlier, as the plasma is essentially molecular, the negative and molecular positive ions dominate the population of charge carriers.

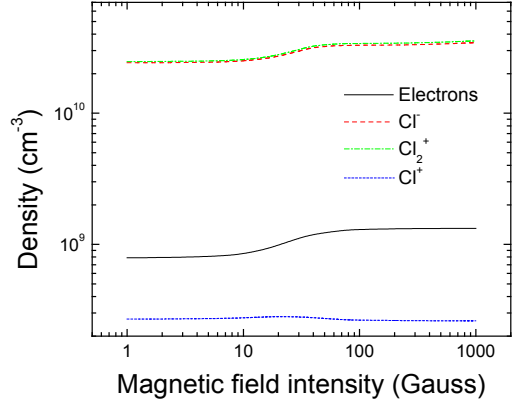

**Figure 4:** Density of charged particles as a function of the magnetic field intensity for an initial gas pressure of 10 mTorr.

The results presented above exhibit some features that were previously observed in pure argon [2,7]. However, as chlorine is a molecular electronegative gas, it is additionally subjected to several kinds of volume losses (in particular ion-ion and dissociative recombination) that may affect the plasma kinetics. In order to estimate the importance of the volume losses and their dependence on plasma conditions, we have calculated the volume loss rates of the various plasma species and well as the creation. Under steady-state conditions, the difference between creation and volume losses is due to diffusion (see Eq. (1)). The results obtained for electrons are presented in Fig. 5 as a function of the magnetic field intensity at a gas pressure of 1 mTorr. Figure 5 shows that losses are diffusion-controlled at low magnetic field while volume losses (mainly dissociative attachment and dissociative recombination) predominate above about 100 G. The decreasing importance of the diffusion contribution to losses as $B_0$ increases is due to the confinement that progressively inhibits the diffusion motion in the direction perpendicular to the magnetic field. On the other hand, one notes a significant increase of volume losses with $B_0$ which is caused by the corresponding increase of $n_e$ as shown in Fig. 3.

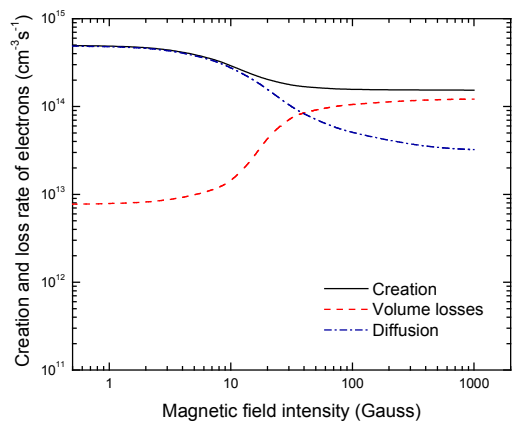

**Figure 5:** Dependence of the volume losses and diffusion on the magnetic field intensity at 1 mTorr. The electron creation rate is shown for comparison.

At 10 mTorr, Fig. 6 shows that volume losses are smaller than diffusion losses but remain comparable to it. In contrast, at higher field, diffusion can completely be ignored. Overall, in this pressure regime, neglecting diffusion cause a significant error, except when a significant magnetic field is applied (> 50-100 G). Similar trends were observed for all the other charged species. Therefore, models in which only volume losses are considered for describing high-density plasmas need to be revisited to the light of the present results.

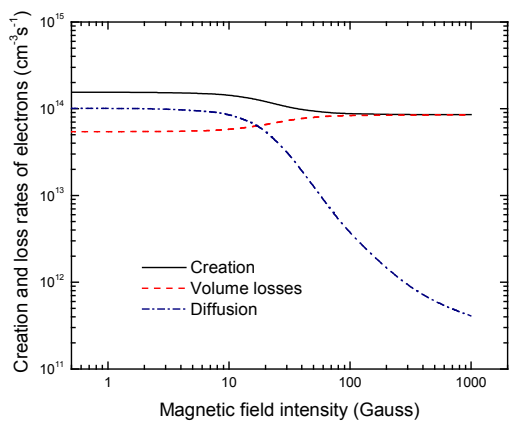

**Figure 6:** Dependence of the volume losses and diffusion on the magnetic field intensity at 10 mTorr. The electron creation rate is shown for comparison.

### 4. Conclusion

We have shown that the model that we have developed reproduces the pressure dependence of the dissociation degree in a chlorine high-density plasma. It was further shown that at sufficiently low pressure, the plasma species depend on the intensity of the confinement magnetic field. At low magnetic field, the concentration of molecular neutral and ions, as well as negative ions is significant, while at higher field, atomic neutral and positive ion species, as well as electrons dominate the plasma. For higher pressure, the magnetic field intensity has only a weak influence on the plasma composition. Finally, we have demonstrated that, in most general conditions, models need to include diffusion to correctly describe the plasma behavior.